%% file: main.tex
\documentclass[sigconf]{acmart}
\usepackage{color}
\usepackage{amssymb}
\usepackage{graphicx}
\usepackage{framed}
\usepackage{amsmath}
\usepackage{url}
\usepackage{multicol}
\usepackage{listings}
\usepackage{xcolor}
\usepackage{array}
\usepackage{float}
\usepackage{graphicx}
\usepackage{subcaption}

\restylefloat{table}
\newcolumntype{P}[1]{>{\centering\arraybackslash}p{#1}}

\colorlet{punct}{red!60!black}
\definecolor{background}{HTML}{EEEEEE}
\definecolor{delim}{RGB}{20,105,176}
\colorlet{numb}{magenta!60!black}
\lstdefinelanguage{json}{
	basicstyle=\normalfont\ttfamily,
	numbers=left,
	numberstyle=\scriptsize,
	stepnumber=1,
	numbersep=8pt,
	showstringspaces=false,
	breaklines=true,
	frame=lines,
	backgroundcolor=\color{background},
	literate=
	*{0}{{{\color{numb}0}}}{1}
	{1}{{{\color{numb}1}}}{1}
	{2}{{{\color{numb}2}}}{1}
	{3}{{{\color{numb}3}}}{1}
	{4}{{{\color{numb}4}}}{1}
	{5}{{{\color{numb}5}}}{1}
	{6}{{{\color{numb}6}}}{1}
	{7}{{{\color{numb}7}}}{1}
	{8}{{{\color{numb}8}}}{1}
	{9}{{{\color{numb}9}}}{1}
	{:}{{{\color{punct}{:}}}}{1}
	{,}{{{\color{punct}{,}}}}{1}
	{\{}{{{\color{delim}{\{}}}}{1}
	{\}}{{{\color{delim}{\}}}}}{1}
	{[}{{{\color{delim}{[}}}}{1}
	{]}{{{\color{delim}{]}}}}{1},
}

\urldef{\mailsa}\path|firstname.lastname@Anonymised.org|

\begin{document}

\copyrightyear{2018}
\acmYear{2018}
\setcopyright{acmcopyright}
\acmConference[JCDL '18]{The 18th ACM/IEEE Joint Conference on Digital Libraries}{June 3--7, 2018}{Fort Worth, TX, USA}
\acmBooktitle{JCDL '18: The 18th ACM/IEEE Joint Conference on Digital Libraries, June 3--7, 2018, Fort Worth, TX, USA}
\acmPrice{15.00}
\acmDOI{10.1145/3197026.3197054}
\acmISBN{978-1-4503-5178-2/18/06}

% first the title is needed
\title{Contextualised Browsing in a Digital Library's Living Lab}

% a short form should be given in case it is too long for the running head

% the name(s) of the author(s) follow(s) next
%
% NB: Chinese authors should write their first names(s) in front of
% their surnames. This ensures that the names appear correctly in
% the running heads and the author index.
%
%\author{Zeljko Carevic, Sascha Sch\"uller, Philipp Mayr}
 
\author{Zeljko Carevic}
\affiliation{
  \institution{GESIS -- Leibniz Institute for the Social Sciences}
 \streetaddress{Unter Sachsenhausen 6-8}
  \city{Cologne} 
  \state{Germany }
}
\email{Zeljko.Carevic@gesis.org} 

\author{Sascha Sch\"uller}
\affiliation{
  \institution{GESIS -- Leibniz Institute for the Social Sciences}
 \streetaddress{Unter Sachsenhausen 6-8}
  \city{Cologne} 
  \state{Germany }
}
\email{Sascha.Schueller@gesis.org} 
\author{Philipp Mayr}
\affiliation{
  \institution{GESIS -- Leibniz Institute for the Social Sciences}
 \streetaddress{Unter Sachsenhausen 6-8}
  \city{Cologne} 
  \state{Germany }
}
\email{Philipp.Mayr@gesis.org}
\author{Norbert Fuhr}
\affiliation{
  \institution{University of Duisburg-Essen}
 \streetaddress{Faculty of Engineering Sciences, Information Systems}
  \city{Duisburg, Germany} 
}
 \email{fuhr@uni-duisburg.de}

%
% (feature abused for this document to repeat the title also on left hand pages)

% the affiliations are given next; don't give your e-mail address
% unless you accept that it will be published

%
% NB: a more complex sample for affiliations and the mapping to the
% corresponding authors can be found in the file "llncs.dem"
% (search for the string "\mainmatter" where a contribution starts).
% "llncs.dem" accompanies the document class "llncs.cls".
%

%\toctitle{Lecture Notes in Computer Science}
%\tocauthor{Authors' Instructions}

\begin{abstract}		
Contextualisation has proven to be effective in tailoring \linebreak search results towards the users' information need. While this is true for a basic query search, the usage of contextual session information during exploratory search especially on the level of browsing has so far been underexposed in research. In this paper, we present two approaches that contextualise browsing on the level of structured metadata in a Digital Library (DL), (1) one variant bases on document similarity and (2) one variant utilises implicit session information, such as queries and different document metadata encountered during the session of a users. 
We evaluate our approaches in a living lab environment using a DL in the social sciences and compare our contextualisation approaches against a non-contextualised approach. For a period of more than three months we analysed 47,444 unique retrieval sessions that contain search activities on the level of browsing. Our results show that a contextualisation of browsing significantly outperforms our baseline in terms of the position of the first clicked item in the result set. The mean rank of the first clicked document (measured as \textit{mean first relevant - MFR}) was 4.52 using a non-contextualised ranking compared to 3.04 when re-ranking the result lists based on similarity to the previously viewed document. Furthermore, we observed that both contextual approaches show a noticeably higher click-through rate. A contextualisation based on document similarity leads to almost twice as many document views compared to the non-contextualised ranking. 
\end{abstract}
\maketitle

\input{sections/introduction}

\input{sections/relatedwork}

\input{sections/approach}
\input{sections/methodology}

\input{sections/results}

\input{sections/discussion}

\input{sections/strengths}
\input{sections/conclusion}
\bibliographystyle{ACM-Reference-Format}
\bibliography{references} 

\end{document}

%% file: sections/introduction.tex
\section{Introduction}

Exploratory search in a DL, especially on the level of browsing, is a frequent strategy when looking for related content  \cite{carevic2016survey}, \cite{carevic2017investigating}, \cite{mayr2017complete}. 
Due to structured metadata that annotate the content of scholarly DLs, users are able to explore the content based on shared characteristics like keywords, classifications, or author information. These paths to start exploratory search in a DL are referred to as \em Stratagems \em \cite{bates1990should} or search stratagems. Although stratagems are encouraging the user for further exploration, the system support on this level is rather low. Modern DLs support stratagem search as simple boolean filters that disregard information about the present user, his or her information need, and session activities. One way to enhance browsing on the level of stratagems is to integrate the users' search context and thus, tailor search results based on previous search activities \cite{Smeaton2005}. Instead of filtering the documents based on shared characteristic as it is the current state-of-the-art in many DLs, we propose a contextualised stratagem search that extends the basic filtering by re-ranking the results with respect to the users' search context. We develop two short-term contextualisation approaches on the level of stratagems for DLs by utilising implicit user feedback. The goal of our contextualisation approach is to tailor search results on the level of stratagems towards the users' current search task. \newline

Integrating the users' search context to personalise search results has provided large benefits during information seeking. 
While this is true for ad hoc search, the usage of contextual session information during exploratory search especially on the level of \em browsing \em has so far been underexposed in research. 

% In this paper we understand browsing in comparison to searching as a search activity that produces a pre-defined group of information items sharing a particular metadate while searching produces ad hoc collections of information that has not been gathered together before \cite{Hearst}.
Following \cite{hearst2009search}, we understand browsing as a search activity that leads to a pre-defined group of information items sharing a particular metadata, while searching produces ad hoc collections of information that have not been gathered together before.

In this paper, we present two short-term contextualisation approaches that integrate the users' search context in exploratory search. A) A contextual feature that re-ranks documents based on their similarity to the previously viewed document and B) a contextual ranking feature that re-ranks documents based on the users' search activities (e.g. the previously entered query terms and viewed documents) in the current retrieval session. \newline
In this study, we aim to answer the following research question: 

\begin{itemize}
	\item Can we improve the \textbf{effectiveness} of exploratory search on the level of browsing by using contextual ranking features in comparison to a non-contextual ranking feature?
	%\item How is the \textbf{usefulness}  of short term contextual ranking features compared to long term contextual ranking features?
\end{itemize}
One of the major challenges in the evaluation of contextualisation approaches is that large scale log data is usually hard to obtain. 
For this reason, we evaluate our approaches using the real life DL's living lab for the social sciences \em Sowiport \em and compare them against a non-contextualised ranking of exploratory search. Using an A/B/C testing, each user is randomly assigned to one approach for the entire session. To evaluate the effectiveness of each approach we measure: a) the position of the first clicked item which we refer to as "mean first relevant" (MFR)  b) the click-through rate and \linebreak c) the usefulness in terms of implicit relevance feedback. \newline

Between September/12/2017 and December/20/2017 we analysed 607,109 sessions of transaction logs provided by our living lab. Our results show that contextualising search activities on the level of browsing significantly improved the retrieval quality in terms of MFR \cite{Fuhr:17b}. \newline

In modern DLs the most common feature to narrow down search results based on certain document features are \em facets \em which are widely implemented nowadays. Facets are described as a "set of meaningful labels organized in such a way as to reflect the concepts relevant to a domain" \cite{hearst2006clustering}. Although empirical studies identified various beneficial aspects of faceted browsing \cite{fagan2010usability}, they usually operate on the level of simple Boolean retrieval. Furthermore, facets require the user to interact with the result lists and to select each filter criterion one after another. In contrast to facets, our contextualisation approach tailors search results on the level of stratagems based on the users' previous interactions without additional effort by the user. In other words: our approach re-ranks documents that result from stratagem search with respect to the users' search context while the advantages of faceted browsing remain unaffected and can still be utilised by the user.\newline

The paper is structured as follows. In the following Section we present the use case of our living lab study. In Section \ref{rel_section} an overview on related work is provided. Our contextualisation approaches and the baseline are presented in Section \ref{appr_section}. In Section \ref{methodology}, we describe the metrics used to evaluate the effectiveness of our approach followed by a presentation of the results in Section \ref{res_section}. In Section \ref{disc_section}, we discuss the results. Insights into the strengths and weaknesses of our living lab study are presented in Section \ref{strength_sec}.

\subsection*{Living Lab Use Case} 
The idea of a living lab is to understand information seeking behaviour in situ involving and integrating users within the research process and providing a context for testing and evaluating of IR models, methods and systems \cite{azzopardi2011towards}. In general, a living lab is an environment in which researchers are facilitated to test their approaches in real life applications with real life users interacting with the system. One example of a living lab evaluation campaign  was initially provided by the CLEF initiative and is now continued as \em TREC - Open Search\footnote{\url{http://trec-open-search.org/}} \em. 

We evaluate our contextual ranking features in  \emph{Sowiport}, a  Digital Library for the social sciences \citep{Hienert2015} that is based on VuFind 2 with an Apache Solr 5.3 index. For the present study we utilised Sowiport as a Living Lab environment in which we injected our re-ranking based on contextualisation. 
Sowiport comprises about 9.7 mio. literature references from 23 different databases covering topics from social and political sciences. 
On a weekly basis, Sowiport reaches around 20,000 unique users. A document in Sowiport usually covers information about the author(s),  keywords, classifications, source information (journal or conference proceedings). All these document properties are implemented as hyperlinks and when employed, a result list of documents that share that particular metadata is generated.

\begin{figure}[h]
	\centering
	\includegraphics[width=1\linewidth]{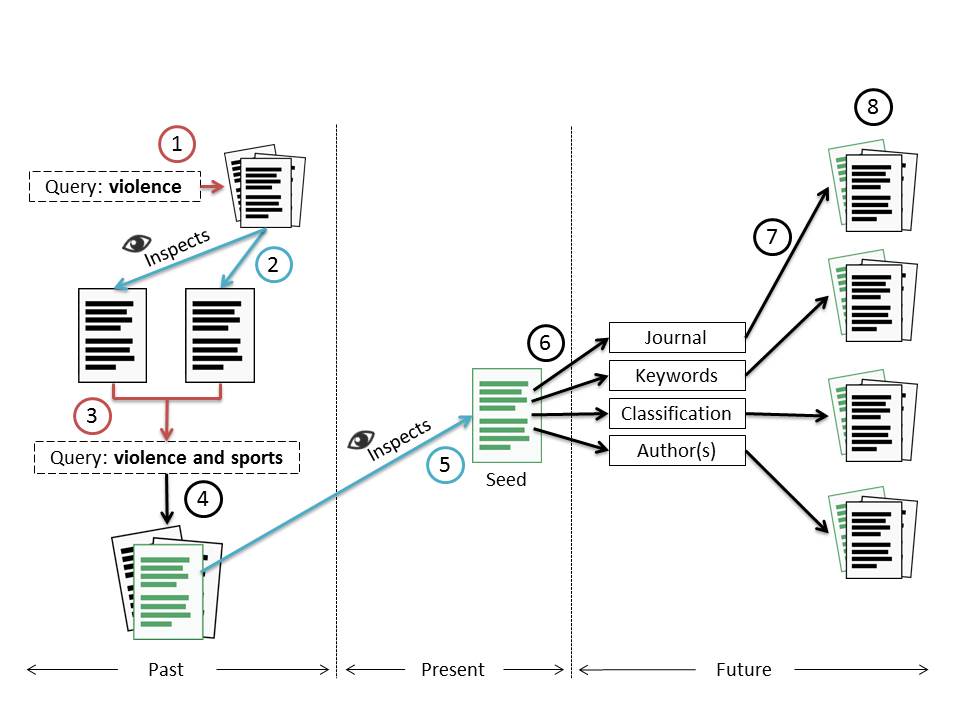}
	\caption{Schematic visualisation of contextualised stratagem searching.} 
	\label{stratagemBasicExample}
\end{figure}

A schematic visualisation of our contextualisation approach can be found in Figure \ref{stratagemBasicExample}. In this example, we have a user who is seeking information on the topic \em violence and sports\em. After entering a query (1), inspecting two documents (2) and refining the query (3), the user has found a document of interest in a result set (4: highlighted green in the figure) that he inspects in detail (5 and 6). To seek further related content the user could for instance look at the journal the document was published in or select a certain keyword that is contained in the current document. Each of these interactions (7) leads to a new result list containing documents that shares the same attribute with the seed document which is also part of the result list (8). Our approach is to re-rank these result lists based on contextual information about the users search sessions. 

\begin{figure*}[h]
	\caption{Example of a keyword search in Sowiport.}
	\centering
	\begin{subfigure}{0.8\linewidth}
		
		\includegraphics[width=1\linewidth]{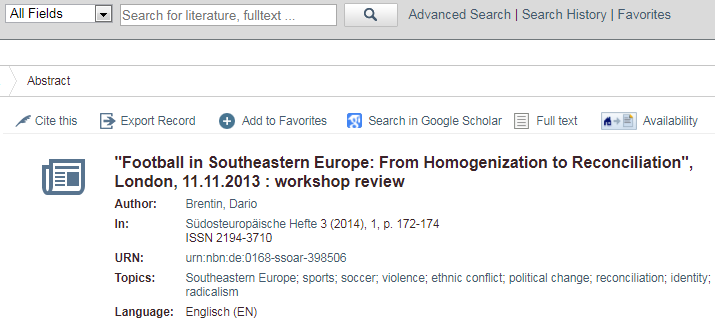}
		\caption{Example document from Sowiport retrieved via a query search for "violence and sports".} 
		\label{basic_document_example}
	\end{subfigure}	\vspace{5mm}%
	
	\begin{subfigure}[b]{0.8\linewidth}
		\includegraphics[width=\linewidth]{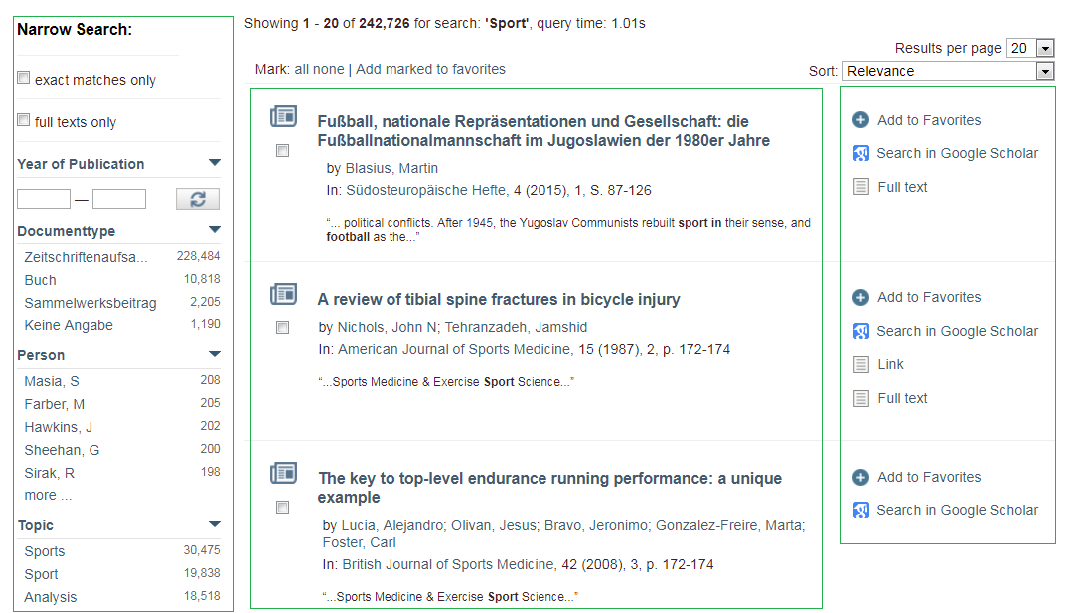}
		\caption{Example result list after a stratagem search for the keyword "sports".}
		\label{similarity_example_res_list}
	\end{subfigure}
	\label{example_fig_strat_search}
\end{figure*} 

A more practical example is displayed in Figure \ref{example_fig_strat_search}a and \ref{example_fig_strat_search}b. In this example a document with the title "\em Football in Southeastern Europe: \em..." has been retrieved via the query for "violence and sports". To perform a stratagem search a user could select a keyword, the name of the author, or the journal (\em S\"udosteurop\"aische Hefte -- Southeastern Europe Magazine\em). Each of these interactions would lead to a result list of documents containing that particular filter criterion. Such a result list is displayed in Figure \ref{example_fig_strat_search}b which was generated by selecting the keyword \em sport \em from the seed document. In this example, the ranking has been contextualised towards the similarity of each document in the result set to the seed document. The displayed result list contains various characteristics worth noticing: One can see that the top ranked document is highly related to football in the former Republic of Yugoslavia which shows a topical relatedness to the seed document. Furthermore, one can see that the result list contains documents from different languages. In this case the top ranked document is in German. A click on a document in the result list leads to a detailed view of that particular document. 
It is important to notice that we did not modify the interface of the result list but rather only re-rank the results towards their similarity to the seed document. Thus, the user is still able to use features like facets or to narrow down the search results based on the year of publication. This is displayed in the left box of Figure  \ref{example_fig_strat_search}b. %todo: numbering of the figure is not ok

For each document a bookmarking feature, a lookup functionality in Google Books and Google Scholar, and a check for the availability in the local library is placed next to each record as well as a link to the full text of that record, if available. This is displayed in the right box of Figure \ref{example_fig_strat_search}b.  %todo: numbering of the figure is not ok

%% file: sections/relatedwork.tex
\section{Related Work}\label{rel_section}
In the early 2000s, Lagoze et al. \cite{lagoze2005} and Smeaton and Callan \cite{Smeaton2005} made a first move to put the concept of \em context \em into the perspective of DL research. They postulate that a DL "should be contextual, expressing the expanding web of inter-relationships and layers of knowledge" \cite{lagoze2005}. In this sense, they called for \em proactive \em \cite{Smeaton2005} or \em collaborative \em \cite{renda2005} or \em adaptive \em \cite{Frias-Martinez:2009} DLs. Since that time, state-of-the-art DLs have not changed much and remain often in the "simple catalog model" \cite{lagoze2005}. Personalisation and recommender systems have evolved and found their way into modern DLs, but deep contextual system support and user guidance are often still missing. \newline

%Some more recent contextualisation approaches can be found in \cite{Coyle2008,Wu:2008}

Exploratory search tasks usually comprise search activities on the level of learning and investigating that go beyond simple look up tasks such as known item search \cite{marchionini2006exploratory}. 
Due to the complexity of exploratory search tasks, various search activities on the level of moves, tactics, and stratagems are typically involved \cite{carevic2017investigating}. 
To date numerous studies have been conducted that aim to understand users' search behaviour and search activities during exploratory search. Ellis \cite{ellis1989behavioural} studied the search behaviour of social scientists and identified six generic features: \em Starting \em  (e.g. to identify a paper to start with), \em Chaining \em(e.g. to follow references in a certain paper), \em Browsing \em(e.g. to browse all papers by a certain author), \em Differentiating \em(e.g. to judge a source based on their nature), \em Monitoring  \em(e.g. to subscribe to an alerting service) and \em Extracting  \em(e.g. to identify material in a well-known journal). \newline

Contextualisation (in web search more commonly referred to as personalisation) has drawn a lot of attention in web search, e.g. (\cite{teevan2011understanding}, \cite{matthijs2011personalizing}, \cite{sepliarskaia2017simple}) and is distinguished between short-term (the current session) and long-term (over many user sessions) personalisation. While short-term personalisation provides better results for recent search interests, it lacks support of the users' long-term interests. Long-term personalisation on the other hand is suitable to capture the users' long-term interests but may not represent recent shifts in search interests \cite{bennett2012modeling}. In this paper, we focus on short-term personalisation which we refer to as contextualisation.

Constructing the user context can be done in various ways like as per explicit (a user explicitly providing feedback information) or implicit feedback which utilises all the information available about the current user, like for instance the viewed/skipped documents or the set of query (re-)formulations. A comparison between implicit and explicit feedback is presented in \cite{white2002comparing}. In total, 16 participants were asked to solve 16 tasks that were taken from the TREC-10 interactive track. In their experiment two interfaces were developed that connect to the Google search engine: one that uses implicit and one that uses explicit relevance feedback. The results showed no significant difference in terms of search effectiveness between the usage of implicit and explicit relevance feedback with regards to viewed result pages, task completion, and task time.

In the  \em TREC session track  \em \cite{Kanuoulas/etal:11} which ran from 2010 to 2014, the main goal was to consider the session history for (re-)ranking the output for subsequent queries. The TREC session track combines a given corpus, topics, queries and relevance judgments with retrieved results, click data and dwell times from crowd workers conducting searches within a session. Similar is the  \em INEX Interactive Track \em \cite{Pharo/etal:11a} where different tasks of real users are conducted (see an overview of different interactive IR data sets in \cite{belkin_PIR-2017}). In contrast to the TREC evaluation setup, we only consider specific queries in the form of contextualised browsing, and we evaluate our approach in a living lab. 

\subsection*{Implicit relevance feedback}
In \cite{white2004simulated} various implicit feedback models are evaluated each aiming to enhance the representation of a user's information need. Each model is constructed by gathering relevance information from the user's exploration of the particular IR system. The different variations are evaluated using searcher simulations. The authors show that each implicit feedback model increased search effectiveness through query expansion. 
In \cite{ruthven2003incorporating} implicit relevance feedback is incorporated for query expansion and to select new terms that are added to the user's query. The authors conducted five experiments on incorporating behavioural information into the process of relevance feedback with 30 participants and 6 tasks. They observed the overall search behaviour, search effectiveness and the subjects perception of the different systems. The results indicate that search behaviour can be utilised in the process of relevance feedback.
In \cite{shen2005context} the authors studied the effect of implicit feedback, including query history and click-through history to improve information retrieval performance. They developed four context-sensitive language models that use statistical language models which incorporate context information into a basic retrieval model. Their experiments showed that the retrieval performance could be substantially improved without requiring any user effort. In \cite{sepliarskaia2017simple} a long-term personalisation approach is developed that incorporated click behaviour into the document ranking. The approach is evaluated with data from the \em Personalized Web Search Challenge \em organised by Yandex and Kagell where researchers are able to provide their own solution to a personalisation task on Yandex log data. White et al. \cite{white2013enhancing} developed a task-based personalisation approach that utilises session history from other users with similar tasks referred to as \em groupisation\em.   
A contextualisation approach aiming to support browsing in a local web site is presented in \cite{alhindi2015profile}. Based on the implicit feedback of groups of users, a document summarisation for browsing is developed. By matching all queries from a group of users a document summarisation was developed  that performs on two levels: single document summarisation and a multi document summarisation, i.e. a summary that is generated from a collection of related documents.

\subsection*{Previous Work}
The present paper is part of a large scale investigation on the usage of stratagems in exploratory search. In a position paper, we discussed the idea of contextual exploratory search \cite{carevic2015extending}. We presented a concept for bibliometric enhanced stratagem searches that contextualises search activities and integrates further re-ranking features like co-citation analysis and author centrality.
In \cite{carevic2016survey} we presented a first approach on gathering a deeper understanding on the usage of stratagems by conducting an online survey with 128 respondents from twelve different fields of research. Our survey showed a general need for a contextual ranking in exploratory search which we tested using a journal run  scenario in which the respondents were asked to arrange the content of a journal run based on two contextual features and four non-contextual features like date or title. The results of the survey showed that the respondents assess the ranking features based on contextualisation noticeably higher than the four non-contextual features. In a recent study \cite{carevic2017investigating}
on exploratory search in a DL, we showed that the majority of search activities were performed on the level of browsing in comparison to queries and the usage of recommendations. A similar observation can be found in \cite{white2007investigating} in which the authors showed that less than one third of the interactions belonged to search result pages while the remainder belongs to pages that lie on the hyperlink trail from a search result page.

%% file: sections/approach.tex
\section{Contextualised Re-Ranking Approach}\label{appr_section} 
We compare three methods for the ranking of exploratory search results. The first approach is the default ranking of Sowiport while the other two approaches are developed to support contextual browsing. We apply contextualised browsing on the level of stratagems and, thus, need a seed document to start with. 
From a set of documents $ D = {\{d_1, . . . , d_n\}}$ the seed document $d_s$ is defined as the document which is currently opened and examined by a user and contains a set 
of possible browsing activities (\em BrowsAct\em)  which are defined as follows: 
\begin{equation*}
\begin{split}
BrowsAct(d_s)=\{Keywords_{d_s},Authors_{d_s},Classifications_{d_s},\\ Journal_{d_s}\}
\end{split}
\end{equation*}

where $ Keywords_{d_s} = {\{k_1, . . , k_j\}}$ is a set of keywords, \linebreak $ Authors_{d_s} = {\{a_1, . . , a_k\}}$ is a set of authors, $ Classifications_{d_s} = {\{c_1, . . , c_l\}}$ is a set of classifications in $d_s$ and  $Journal_{d_s}$  is the journal the seed  document $d_s$ was published in. For all three approaches we excluded the seed document from the result set ($d \neq d_s$) to prevent a potential click bias towards the previously inspected document\footnote{During a pretest, we observed an extraordinary high click rate on the duplicate entry of the result list.}.
%The general ranking function is denoted as:

%\begin{equation}
%R :\{Q\}\rightarrow D
%\end{equation}

%Giving a filter query $Q$  the ranking function $R$ returns a set of relevance scored documents $D$ available in the corpus.
\subsection{Default Ranking} \label{df_ranking}
In this section we describe the default ranking of Sowiport which we use as baseline. The default ranking is based on SOLR and can be formalised as follows: 
\begin{equation}
\label{eq_filter_basic}
\begin{aligned}
DR: Q \times D \rightarrow \mathbb{R}
\end{aligned}
\end{equation}
%Where 
%\begin{equation*}
%T =  \{Keywords_{d_s}, Authors_{d_s},Classifications_{d_s},Journal_{d_s}\}.
%\end{equation*}
A filter query $q$ (selected by the user) is submitted to an expansion function which takes the filter query and expands it with synonyms and corresponding translations (see conceptual model in \cite{mayr2008}). If, for instance, a user clicks on the keyword "violence", this keyword is first expanded with synonyms and different translations. Furthermore, the filter type "keyword" is expanded to related metadata fields. The actual ranking is then provided by boosting the results based on the contained metadata field type. The boosting order is predefined by the DL which assumes that different metadata fields are of higher value than others (e.g. keywords $>$ free keywords). The resulting list of ranked documents in this approach has no connection to the previous search activities of the user and is basically a simple Boolean filter which is extended by a boosting on particular metadata fields. 
We decided to utilise the described Sowiport default ranking as a baseline although an out-of-the-box VuFind-Solr  configuration which performs a simple Boolean filter would decrease the complexity and allow a better reproducibility of our results. The main reason for this decision is that we want to compare our contextualisation approach to a realistic real life DL ranking which also provides us with a strong and established baseline. 

A simplified example query for the default ranking is displayed in Listing \ref{Example_Default_ranking}. Again, we utilise our fictitious example of a user who is looking at a document about "Football in Southeastern Europe". In this example, the user selected the keyword "violence". One can see that the baseline approach not only performs a Boolean query on the keyword level but extends the query to related fields like in this case the metadata field "keyword\_free" which is an alternative and less formal descriptor for keywords. Furthermore, one can see that the term violence is translated into the German word "Gewalt". The ranking of the retrieved documents is based on TF*IDF whereby the weighting of the fields is taken into account. In our example it can be seen that the metadata field "keyword" is boosted by a higher factor than the "keywords\_free" field.
\begin{lstlisting}[basicstyle=tiny,language=json,caption={Example query for the default  ranking},label=Example_Default_ranking]
q => keyword:((violence OR "Gewalt")^400 OR
keyword_free:((violence OR "Gewalt"))^250
\end{lstlisting}

In the following two sections, we describe our contextualisation approach which is then compared against the baseline. 

\subsection{Re-Ranking based on Document Similarity}
In this approach we perform a re-ranking of the result list based on the similarity of each document in the result set compared to the seed document. \newline

This is described formally as follows: 

\begin{equation}
\label{eq_filter_sim}
\begin{aligned}
SR: Q \times D \times D_s \rightarrow \mathbb{R}
\end{aligned}
\end{equation}
%\begin{equation*}
%R_2=\{d \in D \mid d \neq d_s,d \in R(exp_2(fq,T,d_s))\}
%\end{equation*}

The ranking function $SR$ is an extension to the default ranking $DR$. The ranking function takes the filter query and the seed document ($D_s$) as input to produce a ranked list of documents. 
In a first step, the documents are filtered correspondingly to the default ranking. Instead of only boosting the documents based on field types containing the filter query, the ranking is extended by a similarity score of each document compared to the seed document. To estimate the similarity of each document compared to the seed document we utilise the \em MoreLikeThis (MLT) \em query parser\footnote{\url{http://archive.apache.org/dist/lucene/solr/ref-guide/apache-solr-ref-guide-5.3.pdf}} built in SOLR  
which is usually employed to provide related documents to a given seed document. An example query for the re-ranking based on document similarity is provided in Listing \ref{Example_Similarity_ranking}.  One can see that the query is based on the default ranking described in Section \ref{df_ranking} but uses the \em MLT \em  query parser. The seed document is specified by the \em DocID \em parameter. To compute the similarity of all documents to the seed document  we use the keywords, journal information, the abstract (in different languages if available), and the author names of the seed document.  These are specified in the \em qf \em parameter in Listing \ref{Example_Similarity_ranking}.

\begin{lstlisting}[basicstyle=tiny,language=json,caption={Example query for the re-ranking based on document similarity},label=Example_Similarity_ranking]
q => {!mlt}DocID
keyword:((violence OR "Gewalt")^400 OR
keyword_free:((violence OR "Gewalt"))^250
qf=authors,keywords,journal,abstract
\end{lstlisting}
%In most cases the top ranked document in the result list is equal to the seed document as it has the highest similarity  score\footnote{This is not always the case as \em %anonymised portal \em contains duplicate documents and thus there may be multiple versions of a certain document.}. During a first pretest, we observed an extraordinary high click %rate on the duplicate entry of the result list. We assume that the users confused this document with a newly discovered one and thus were biased towards inspecting the same %document again. As a consequence we decided to exclude the seed document from the result lists. We did this for all three approaches to balance the results.\footnote{One %shortcoming of this decision is that the rare case may appear in which the result list does not contain any documents as the stratagem would lead only to one result which is equal to %the seed document and thus be removed.}

\subsection{Re-Ranking based on Session Details}
In this approach, the re-ranking is performed based on the user's session context which is derived from a set of actions the user has performed  during the current search session. \newline This is described formally as follows: 
\begin{equation}
\label{eq_filter_sim}
\begin{aligned}
CR: Q \times D \times U_c \rightarrow \mathbb{R} %todo: check U_c or U_s??
\end{aligned}
\end{equation}
%\begin{equation*}
%R_3=\{d \in D \mid d \neq d_s,d \in R(exp_3(q,T,U_c))\}
%\end{equation*}
where 
\begin{equation*}
U_c = \{ Keywords(U),Categories(U),Queries(U)\}
\end{equation*}
The ranking function $CR$ is again an extension to the default ranking $DR$. %? should be DR ?
The function takes the filter query and the user's search context  ($U_c$) as input to produce a ranked list of documents. Again, the first step is to filter all documents corresponding to the default ranking. The re-ranking based on the user's search activities  is then provided by boosting the filtered documents using the user's session context ($U_c$) which consists of three features: 
keywords, categories and the queries a user $U$ has submitted. 

We create the keyword and category features from two sources: a) each keyword/category is considered that was contained in the list of documents the user has seen during the session, b) each keyword/category is considered that was contained in documents within a result set during the session. We then sort the list of keywords and categories based on the number of their occurrences within the session. If a certain keyword has appeared five times in the list of viewed documents and four times in documents within the result list, the score for that particular keyword is nine. To reduce noise, we limit the number of keywords and categories to the top three for each feature and normalise the count to range between zero and one. Regarding the queries, we did not create a threshold as we assume that all queries are equally important to describe the user's information need and do not contain any noise. 
An example session context could be described as shown in Listing \ref{Example_Session_Context}: 

\begin{lstlisting}[basicstyle=tiny,language=json,caption={Example session context},label=Example_Session_Context]
{"query":"violence sports","rank":1},
{"keyword":"Football","rank":1},
{"keyword":"Radicalism","rank":0.5},
{"keyword":"Ethnic Conflict","rank":0.5},
{"category":"Political Sociology","rank":1},
{"category":"Decision Making","rank":0.66},
{"category":"Sociology","rank":0.66}}
\end{lstlisting}
In this example, the user has submitted the query \em violence sports\em. From the corresponding result set and the viewed document, the categories and keywords are derived with a rank that represents the normalised frequency of that term in the result list and the viewed document. It may happen that we have an overall occurrence count of 1 resulting in a rank of 1 for each keyword and classification, which is the case when a user enters our DL from a web search engine and goes straight to the detailed view of a document. In this case, we use each keyword and classification from the seed document as session context. 

The actual boosting is performed in the following order: by the entered query terms, the keywords, and the classifications each multiplied by their normalised rank (see Listing \ref{Example_Session_Context}).

An example query for the re-ranking based on session details is displayed in Listing \ref{Example_Session_details_ranking}. The session context in this example is derived from the example session context in Listing \ref{Example_Session_Context}. The first part of the query is based on the default ranking which is extended by a boosting parameter \em bq \em. In line 5, it can be seen that we look for the previously entered query term \em violence sports \em in the title field which is also the metadata field with the highest boosting value. In the lines 6 to 8 we boost documents that contain the most frequent keywords in the session context and boost each keyword with a decreasing factor depending on the keywords' frequency in the session context. In line 9 to 11 we boost documents that contain the category terms from the session context.
\begin{lstlisting}[basicstyle=tiny,language=json,caption={Example query for the re-ranking based on session details},label=Example_Session_details_ranking]
q => 
keyword:((violence OR "Gewalt")^400 OR
keyword_free:((violence OR "Gewalt"))^250
[bq]= 
(title:violence sports^1700) OR
(keyword:Football^1200 OR
 keyword:Radicalism^1080 OR 
 keyword:Ethnic Conflict^1080) OR
(category:Political Sociology^800 OR
 category:Decision Making^560 OR 
 category:Sociology^560 )
\end{lstlisting}

%Using the above session context, we again re-rank the result lists produced from a stratagem search in two steps: First, we filter the documents in the same manner as the default behaviour of our system. In a second step, we use the above session context to boost the results from the filtering process in step one. The boosting is done in the following order: by the entered query terms, the keywords and the classifications each multiplied by their normalised rank (see Listing \ref{Example_Session_Context}). 

%Our goal is to determine the users search interest by inspecting each query, retrieved and viewed document within the users search session.

%% file: sections/methodology.tex
\section{Methodology} \label{methodology}
At the beginning of a session each user is assigned one ranking  method for the entire duration of a session. 
As most users click only on one suggested document, we use the rank position of the first clicked document as quality criterion. The most obvious metric to be used for this case would be mean reciprocal rank, but it has been argued in \cite{Fuhr:17b} that this metric is not on an interval scale, and thus the mean cannot be applied. Instead, we used the proposed alternative \em mean first relevant (MFR) \em which takes the rank position of the first clicked document in a result set and computes the arithmetic mean for all result sets that were generated using a stratagem search.  This measure is proportional to the effort a user has to invest in finding the first relevant document in a result list: An MFR value of \em x \em represents the \em x\em-fold effort in comparison to the ideal value of 1. As the number of documents in a result set varies, it may happen that a result set only contains a small number of documents. One typical example is the search for other documents of a certain author or the search for a highly specific keyword. In this case, the MFR is usually rather low and might bias our results. Therefore, we additionally measured the MFR$\geq$20 for all result sets that contain at least 20 documents which is the default number of documents on the first result page in Sowiport. Due to potential outliers that may distort our results, we furthermore disregard document views that were performed on the third page of a result set (first relevant $>$ 40) when measuring MFR.
\newline

Besides the MFR values, we measure other session-related features like session length, number of document views, session duration, and usefulness of stratagems. To measure the usefulness of stratagems, we apply a measure previously described in \cite{hienert2016usefulness} in which different interaction signals within a session can be used to estimate the success of a search session.
For instance, if a user bookmarks a certain document one can consider this document as relevant. The list of implicit relevance signals considered to measure the usefulness is displayed in Table \ref{implicit_relevance_table}.

	\begin{table}[H]
\caption{List of implicit relevance feedback signals}
	\label{implicit_relevance_table}
	\begin{tabular}{ |l|p{5cm}| } 
		\hline
		Short & Description\\\hline
		Add to favourites & Bookmark a single or multiple records to favourites\\
		Goto Google Scholar & Search a record in Google Scholar \\
		Goto Google Books & Search a record in Google Books\\
		Goto Fulltext & View the full text of the record\\
		Goto local availability & Check availability in the local library\\
		Export record & Export the record in different citation styles or via e-mail\\
		
		\hline

	\end{tabular}
	\end{table}

To estimate if and how useful a stratagem was for the outcome of a search session, we measure the usefulness by observing the number of positive signals in the log file after the  usage of a stratagem on two levels:
\begin{itemize}
\item Local usefulness \\
The local usefulness counts the total number of implicit relevance signals on a document that was contained in the result set immediately after a stratagem run. For a visual example of implicit relevance feedback on a result set see the right box in Figure \ref{example_fig_strat_search}b.
\item Global usefulness \\
For each session, we count the total number of implicit relevance signals contained in the entire session after the first usage of a stratagem. 
\end{itemize}

Using the local usefulness we determine the "immediate" relevance of a document in a result set after a stratagem search. The global usefulness on the other hand measures the usefulness of a stratagem search for the entire session.  
%\begin{center}
%$ \displaystyle \text{usefulness} = \frac{1}{\mid\text{Sessions}\mid} %\sum_{n=1}^{\mid\text{Sessions}\mid} count_r(Session_k) $
%\end{center}

%For this work we ignore session boundaries and therefore, we might miss potential shifts in the users search interests during the session and assume that the users search task does not vary during the session. 

%% file: sections/results.tex
\section{Results}\label{res_section}
In total, we analysed 607,109 unique sessions during our living lab study. \newline

\begin{table*}
	\caption{Descriptive statistics for the period of the study}
	\label{descriptive_statistics}
	\centering	
	\begin{tabular}{ |l|P{1.7cm}|P{3cm}|P{2.5cm}|P{1.5cm}|} 
		\hline
		Approach & Total stratagem usage & Document views from stratagem search& Mean interactions per session&Mean dwell time (s)\\\hline
		(A) Baseline & 25,426 & 1,985 &7.61&123.79\\		
		(B) Similarity & 25,475 & 3,212 &7.91&134.98\\
		(C) Session Context & 26,135 & 2,627&7.76&123.27\\						
		\hline			
	\end{tabular}
\end{table*}

Descriptive statistics on the overall usage of stratagems can be found in Table~\ref{descriptive_statistics}. During our experiment, all three approaches were nearly equally distributed among the users (see column 2). Users applied a stratagem search 77,036 times from which 5,839 documents were clicked from the result lists. In column 5, we measured the mean dwell time starting from the first usage of a stratagem until the end of the session. When looking at the dwell time, we identified numerous outliers. We therefore removed all sessions that exceeded a dwell time of more than 20 minutes when measuring the average dwell time. On average, the users continued their search for 2.1 minutes after the first stratagem usage. All three approaches show a similar dwell time. Users that were assigned approach A or C continued their search for 123 seconds while approach B was continued for 133 seconds. In column 4, the mean number of interactions for all sessions containing a stratagem search is displayed. It can be seen that the number of interactions do not differ between the three approaches.  In column 3, the number of document views from stratagem usage per approach is displayed. Both contextualised approaches have a considerably higher number of document views from stratagem search in comparison to the non-contextualised approach. Using the baseline, only 1,985 documents were viewed  while the contextualisation based on similarity received more than 3,200 document views from stratagem searches. The contextualisation based on the session context also clearly outperforms the baseline with 2,627 document views. This tendency can also be seen in the click-through rates which are discussed later on in this section. \newline

To estimate the effectiveness of the contextual ranking features in comparison to the non-contextualised baseline, we utilised the mean first relevant metric (MFR) as described in Section \ref{methodology}: the idea of the MFR metric is to take the position of the first clicked document in the result set.
In Table \ref{mean_first_relevant_total}, the MFR values for all sessions are displayed. The contextualisation based on similarity to the seed document performed best with an MFR of 3.10. The contextualisation based on session details received an MFR of 3.62 while the non-contextualised ranking performed the worst with an MFR of 4.66\footnote{In addition to MFR, we calculated mean reciprocal rank (MRR) values for all three approaches. MRR showed the same tendency as MFR, so we decided not to report MRR separately in this paper.}. Besides the better results in terms of MFR, we can observe a higher click-through rate (denoted as N) for both contextualised approaches. For the contextualisation based on similarity, we observed nearly twice as many first relevant clicks (N=1999) compared to the baseline (N=1078). \newline

The contextualisation based on the session context also clearly outperforms the baseline with 1,571 first relevant clicks.

\begin{table}
	\centering
	\caption{Mean first relevant}
	\label{mean_first_relevant_total}	
	\begin{tabular}{ |l|P{1.5cm}|c|P{1.5cm}|c|} 
		\hline
		Approach & MFR&SD &MFR $\geq$20 & SD \\\hline
		(A) Baseline& 4.66 (N=1078)&6.45& 6.47 (N=607)&7.74\\\hline
		(B) Similarity   &3.10* (N=1999)&4.41&3.39* (N=1528)&4.81\\\hline
		(C) Session context &3.62* (N=1571)&4.74&4.30* (N=1097)&5.33\\\hline		
	\end{tabular}
\end{table}

Due to highly skewed data, we utilised a non-parametric Mann-Whitney U-Test to seek for significant differences in the results with a Bonferroni corrected $p^*=0.016$. We see that both contextualised ranking features significantly outperform the non-contextualised ranking with A $>$ B, $p^*$=0.001 r=0.13 and A $>$ C, $p^*$=0.014 r=0.04. Furthermore, we see that the contextualisation based on similarity significantly outperforms the overall session context with (B $>$ C, $p^*$=0.001 r=0.04).

As we may have different result set sizes, we furthermore measure the MFR$\geq$20 for all result sets that contain at least 20 documents. The MFR$\geq$20 values are displayed in column 4 of Table \ref{mean_first_relevant_total}. The MFR$\geq$20 values for all three approaches increased with a stronger impact on approach A and C. We can observe that the MFR for the non-contextualised approach increased from 4.66 to 6.47. However, we can not observe a substantial difference when comparing MFR and MFR$>$20. The contextualisation based on document similarity still performs best while both contextualisation approaches still outperform the baseline significantly.
\newline

The efficiency of approach C highly depends on how rich the interaction of the user was before performing a stratagem search. If a user just entered the DL without any previous interaction we have a cold start problem. Therefore, we compared the MFR for different history sizes (defined as the number of interactions prior to the stratagem usage) to evaluate whether the usage of a stratagem later on in a session has any influence on the performance of the approaches. In Table \ref{mean_first_relevant_per_window_size}, the MFR values for 3 different history sizes are displayed.

\begin{table*}
	\centering
	\caption{Mean first relevant for different history sizes (H)}
	\label{mean_first_relevant_per_window_size}	
	\begin{tabular}{ |l|P{4cm}|P{4cm}|P{4cm}|} 
		\hline
		Approach & MFR  $\text{H}\in[2,5]$& MFR $\text{H}\in[6,10]$& MFR $\text{H}\in[11,\infty]$ \\\hline
		(A) Baseline& 4.52 (N=802) & 5.06 (+11.94\%, N=276) &5.49 (+21.46\%, N=112) \\ \hline
		(B) Similarity  & 3.04 (N=1491) & 3.29 (+8.22\%, N=508)&3.59 (+18.09\%, N=215) \\\hline
		(C) Session context  & 3.58 (N=1201) & 3.77 (+5.30\%, N=370)&3.80 (+6.14\%, N=160) \\\hline		
	\end{tabular}
\end{table*}

The results in Table \ref{mean_first_relevant_per_window_size} show that with a growing history size the MFR values of all three approaches increase. Looking at the non-contextualised approach the MFR at early stages of the session is 4.52 while increasing to 5.06 after a history  size $>$ 5. If we compare sessions with a history size $\geq 10$, we can observe that the MFR for all approaches further increases (A=5.49, B=3.59, C=3.80). The differences between the MFR values get even more evident if we look at the percentage increase between the different history sizes. For a history size of $\text{H}\in[11,\infty]$ the MFR values of the baseline increased by 21.46\% compared to a history size of $\text{H}\in[2,5]$ while the contextualisation based on session context only increased by 6.14\%.   However, the number of sessions with a history size $\geq 10$ is comparably low with only 487 sessions. Having a larger sample size would improve the reliability of these observations. \newline

To measure the usefulness of the contextualisation in comparison to the baseline, we utilise implicit relevance signals which can be found in Table \ref{implicit_relevance_table}. In general, we measure the usefulness by observing the number of positive signals in the log file after the first usage of a stratagem on two levels: a) the local usefulness b) the global usefulness.

To exclude outliers from this experiment, we only considered sessions with a total of implicit relevance signals $\leq$ 10 . The results of this experiment are displayed in Table \ref{usefulness_results}.

\begin{table}[H]
	\caption{Usefulness of stratagem browsing per session}
	\centering	
	\label{usefulness_results}	
	\begin{tabular}{ |l|P{1.7cm}|P{1.7cm}|} 
		\hline
		Approach & Local usefulness & Global usefulness\\\hline
		(A) Baseline &  232 &5,385\\		
		(B) Similarity & 628 & 5,684 \\% (N=2906)
		(C) Session Context &  334 & 5,294\\%(N=2677)		
		\hline			
	\end{tabular}
\end{table}
Regarding the number of local usefulness signals displayed in column 2, we can observe that the contextual ranking (B=628,C=334) again outperforms the non-contextualised baseline (A=232). Regarding the global session usefulness, the results show only marginal differences between the contextual approaches and the non-contextual baseline. The similarity approach again performs best with 5,684 implicit relevance signals in total after the first usage of a stratagem search. However, the results for the global usefulness show only marginal differences between the contextual approaches and the non-contextual baseline.

%% file: sections/discussion.tex
\section{Discussion}\label{disc_section}
By measuring the position of the first relevant document in the result set (MFR), we showed that both contextualisation approaches lead to significantly better results in comparison to the baseline. Besides the better results of the MFR we see a considerably higher click-through-rate of the contextualisation (B=1999, C=1571) compared to the baseline (A=1078). Both, the MFR and the click-through rates are strong indicators for a better effectiveness of the contextualised approaches in comparison to the non-contextualised baseline.

When limiting the MFR measure to all result sets that contain at least 20 documents we observed an increase of the MFR for all three approaches. The non-contextualised baseline for instance increased from an MFR of 4.66 to 6.47. 

The results for MFR and the higher click-through rate do not include information about the relevance of the clicked documents but rather about a topical relatedness to the users search interests.

To overcome this problem and to get an insight into the perceived relevance, we additionally utilised the usefulness metric that takes implicit relevance signals into account. This analysis was carried out on two levels: a) the local usefulness of a search result after a stratagem run and b) the global usefulness which measures the implicit relevance signals for the entire session. The results for the local usefulness are in line with the results of the MFR. Both contextualised approaches outperform the baseline. The contextualisation based on document similarity gathered more than twice as many direct relevance signals as the baseline while the contextualisation based on the session context was also considerably higher (A=232,B=628,C=334). This underpins the assumption that the contextualised approaches provide a more effective topical relatedness to the users search interests. By utilising the global usefulness, we measured the number of implicit relevance signals after the first stratagem usage for the entire session. For the global usefulness, we identified only minor differences in the overall success of the sessions. This may be an indicator for a rather equal performance of the three approaches in terms of satisfying the users information need. \newline

For the session context approach we identified a typical cold start problem. A majority of users visit Sowiport via a web search engine which indexes the detail view of a document. Users employing a stratagem for further exploration are very common in the transaction logs. However, in this case we have no information about the users information need and therefore, only little information can be used for contextualisation. The more interactions a user performs the more detailed the session context of the user can be modeled. This was also indicated by the results of our segmentation into different sets of window sizes in Table \ref{mean_first_relevant_per_window_size}. We observed an increase in terms of MFR for larger history sizes. We assume that the performance of the session context depends on the complexity of the search task. If a session contains various interactions this is also an indication for the complexity of a search task. We assume that sessions with a higher number of interactions also have greater demands on the quality of a result list as all three approaches performed worse when the number of interactions increased.

We observed a general tendency for an improved performance of the session context with growing history sizes. The MFR values for the contextualisation based on similarity increased by 18.09\% for larger history sizes in comparison to a history size of $H \in[2.5]$. The contextualisation based on session details only increased by 6.14\%. The non-contextualised approach had the largest increase with 21.46\%. 
Unfortunately, the number of sessions with a history size $\geq 10$ is comparably low with only 487 sessions. Having a larger sample size would improve the reliability of these observations. \newline

To overcome the cold start problem a hybrid approach could be implemented that contextualises browsing based on document similarity at the beginning of a session while longer sessions are contextualised based on session details. As the majority of search sessions in \em Sowiport \em comprises (1 to 5) interactions we can not generalise the results of our session context approach (C). We assume that for highly interactive search sessions the results of the re-ranking will differ and lead to a better performance.

Surprisingly, other session related measures like the dwell time or the number of interactions did not differ substantially between the approaches.

%% file: sections/strengths.tex
\section{Strengths and Limitations of the study}\label{strength_sec}
The results of the transaction log study provide insights towards a general usefulness of contextualised ranking features in exploratory search. Both, short-term and long-term ranking features outperform the baseline. While the baseline describes the general behaviour of many DLs it may not be suitable for comparison with more sophisticated approaches like the ones that are presented in this paper. \newline

One downside of the present paper lies in the nature of transaction log studies. Our results indicate a significantly better ranking in terms of MFR on the level of contextual ranking features. However, we are not able to interpret our results in the light of quality and overall relevance. A similar argumentation is provided in \cite{joachims2003evaluating} where a presentation bias in click-through data is observed.
To gain a more qualitative view on the usage of contextual ranking features a user study could be conducted that gives insights into the relevance of documents and particular reasons for regarding or disregarding certain documents. \newline
We used implicit relevance ratings to measure the overall success of each session. Although the implicit relevance ratings had previously been used to measure the usefulness of DL features \cite{hienert2016usefulness}, they are not evaluated entirely. Thus, the results for this observation are influenced by our subjective relevance ratings of search activities. However, they provide a suitable measure to estimate the usefulness of search results. 

One downside of the present approach is that the user is not aware of the contextual ranking features. In a first attempt, we provided the users with an information box which aimed to give some information on the ranking mechanism and the ability to disable the contextualised ranking. After running a pretest we decided to remove all information about the ranking features on the user interface as this led to confusion and distorted results. 

By assigning each user one approach for the entire session, we kept the overall usage balanced. In this way, we eliminated seasonal effects like the start of a university semester or weekends. On the other hand, we ignored potential shifts in the users' search interest during the session. A more sophisticated approach would be to identify session boundaries. \newline

Another advantage of the present study is that it can be easily reproduced by other researchers and DL developers. As we relied on the MLT query parser which is a standard similarity function build in SOLR one can reproduce our approach in its own environment. The re-ranking based on the session context uses SOLR boost functions which also can be reproduced as long as a suitable user model can be derived.

%However, the influence of the implicit ratings is distributed through all approaches equally and therefore only  
%A higher might provide more reliable results for ncdg. However, the 

%% file: sections/conclusion.tex
\section{Conclusion}
In this paper, we have introduced two contextualisation approaches on the level of browsing in a DL: a) a contextualisation based on document similarity and b) a contextualisation based on implicit feedback about the users' search activities. Using a living lab environment, we evaluated our experiment on 47,444 sessions that contained browsing activities. Our results show that a contextualisation of browsing in DLs significantly outperforms the baseline in terms of MFR. Furthermore, we observed an increase in the click-through rate for our contextualisation approaches in comparison to the baseline. Other session related factors like dwell time or the total number of interactions per session did not differ between the approaches. As we utilised basic SOLR functions like the \em More Like This \em query parser to contextualise stratagem searches, our approach can easily be reproduced by other researchers and DL developers.

Although our contextualisation approach outperforms the baseline, a further investigation is necessary as our living lab study can not give insights into the overall relevance of the viewed documents. To tackle this problem we measured the local and global usefulness which take implicit relevance feedback into account. The results of our local usefulness analysis show that the contextualisation had a strong impact in terms of topical relatedness when selecting a document from the result set. Again, both contextualised approaches outperform the non-contextualised baseline. The global usefulness for the entire session, however, did not show significant differences between the contextualised and the non-contextualised re-ranking. Thus, we can state that the contextualised ranking features performed significantly better in terms of MFR, click-through rate and local usefulness. However, insights on the usefulness of our feature with regards to the users information need can not be derived from our living lab study. The increase in terms of the click-through rate and the significantly better MFR values, however, provide us with a strong indication that users could benefit from contextualised ranking features on the level of stratagems and that further research in this area could be of benefit for the DL community.  
Future work will be to conduct a user study that provides us with qualitative feedback on the effectiveness of our contextualisation approaches. 
 %In comparison to other contextualisation approaches we apply our method on the level of stratagems.

%% file: main.bbl
%%% -*-BibTeX-*-
%%% Do NOT edit. File created by BibTeX with style
%%% ACM-Reference-Format-Journals [18-Jan-2012].

\begin{thebibliography}{34}

%%% ====================================================================
%%% NOTE TO THE USER: you can override these defaults by providing
%%% customized versions of any of these macros before the \bibliography
%%% command.  Each of them MUST provide its own final punctuation,
%%% except for \shownote{}, \showDOI{}, and \showURL{}.  The latter two
%%% do not use final punctuation, in order to avoid confusing it with
%%% the Web address.
%%%
%%% To suppress output of a particular field, define its macro to expand
%%% to an empty string, or better, \unskip, like this:
%%%
%%% \newcommand{\showDOI}[1]{\unskip}   % LaTeX syntax
%%%
%%% \def \showDOI #1{\unskip}           % plain TeX syntax
%%%
%%% ====================================================================

\ifx \showCODEN    \undefined \def \showCODEN     #1{\unskip}     \fi
\ifx \showDOI      \undefined \def \showDOI       #1{#1}\fi
\ifx \showISBNx    \undefined \def \showISBNx     #1{\unskip}     \fi
\ifx \showISBNxiii \undefined \def \showISBNxiii  #1{\unskip}     \fi
\ifx \showISSN     \undefined \def \showISSN      #1{\unskip}     \fi
\ifx \showLCCN     \undefined \def \showLCCN      #1{\unskip}     \fi
\ifx \shownote     \undefined \def \shownote      #1{#1}          \fi
\ifx \showarticletitle \undefined \def \showarticletitle #1{#1}   \fi
\ifx \showURL      \undefined \def \showURL       {\relax}        \fi
% The following commands are used for tagged output and should be
% invisible to TeX
\providecommand\bibfield[2]{#2}
\providecommand\bibinfo[2]{#2}
\providecommand\natexlab[1]{#1}
\providecommand\showeprint[2][]{arXiv:#2}

\bibitem[\protect\citeauthoryear{Alhindi, Kruschwitz, Fox, Albakour,
  et~al\mbox{.}}{Alhindi et~al\mbox{.}}{2015}]%
        {alhindi2015profile}
\bibfield{author}{\bibinfo{person}{Azhar Alhindi}, \bibinfo{person}{Udo
  Kruschwitz}, \bibinfo{person}{Chris Fox}, \bibinfo{person}{M Albakour},
  {et~al\mbox{.}}} \bibinfo{year}{2015}\natexlab{}.
\newblock \showarticletitle{Profile-based summarisation for web site
  navigation}.
\newblock \bibinfo{journal}{\emph{ACM Transactions on Information Systems
  (TOIS)}} \bibinfo{volume}{33}, \bibinfo{number}{1} (\bibinfo{year}{2015}),
  \bibinfo{pages}{4}.
\newblock


\bibitem[\protect\citeauthoryear{Azzopardi and Balog}{Azzopardi and
  Balog}{2011}]%
        {azzopardi2011towards}
\bibfield{author}{\bibinfo{person}{Leif Azzopardi} {and}
  \bibinfo{person}{Krisztian Balog}.} \bibinfo{year}{2011}\natexlab{}.
\newblock \showarticletitle{Towards a living lab for information retrieval
  research and development}. In \bibinfo{booktitle}{\emph{International
  Conference of the Cross-Language Evaluation Forum for European Languages}}.
  Springer, \bibinfo{pages}{26--37}.
\newblock


\bibitem[\protect\citeauthoryear{Bates}{Bates}{1990}]%
        {bates1990should}
\bibfield{author}{\bibinfo{person}{Marcia~J Bates}.}
  \bibinfo{year}{1990}\natexlab{}.
\newblock \showarticletitle{Where should the person stop and the information
  search interface start?}
\newblock \bibinfo{journal}{\emph{Information Processing \& Management}}
  \bibinfo{volume}{26}, \bibinfo{number}{5} (\bibinfo{year}{1990}),
  \bibinfo{pages}{575--591}.
\newblock


\bibitem[\protect\citeauthoryear{Belkin, Hienert, Mayr, and Shah}{Belkin
  et~al\mbox{.}}{2017}]%
        {belkin_PIR-2017}
\bibfield{author}{\bibinfo{person}{Nicholas~J. Belkin}, \bibinfo{person}{Daniel
  Hienert}, \bibinfo{person}{Philipp Mayr}, {and} \bibinfo{person}{Chirag
  Shah}.} \bibinfo{year}{2017}\natexlab{}.
\newblock \showarticletitle{Data {Requirements} for {Evaluation} of
  {Personalization} of {Information} {Retrieval} - {A} {Position} {Paper}}. In
  \bibinfo{booktitle}{\emph{Working {Notes} of {CLEF} 2017 - {Conference} and
  {Labs} of the {Evaluation} {Forum}}}. \bibinfo{publisher}{CEUR-WS.org},
  \bibinfo{address}{Dublin, Ireland}.
\newblock
\urldef\tempurl%
\url{http://ceur-ws.org/Vol-1866/paper_193.pdf}
\showURL{%
\tempurl}


\bibitem[\protect\citeauthoryear{Bennett, White, Chu, Dumais, Bailey, Borisyuk,
  and Cui}{Bennett et~al\mbox{.}}{2012}]%
        {bennett2012modeling}
\bibfield{author}{\bibinfo{person}{Paul~N Bennett}, \bibinfo{person}{Ryen~W
  White}, \bibinfo{person}{Wei Chu}, \bibinfo{person}{Susan~T Dumais},
  \bibinfo{person}{Peter Bailey}, \bibinfo{person}{Fedor Borisyuk}, {and}
  \bibinfo{person}{Xiaoyuan Cui}.} \bibinfo{year}{2012}\natexlab{}.
\newblock \showarticletitle{Modeling the impact of short-and long-term behavior
  on search personalization}. In \bibinfo{booktitle}{\emph{35th SIGIR
  conference}}. ACM, \bibinfo{pages}{185--194}.
\newblock


\bibitem[\protect\citeauthoryear{Carevic, Lusky, van Hoek, and Mayr}{Carevic
  et~al\mbox{.}}{2017}]%
        {carevic2017investigating}
\bibfield{author}{\bibinfo{person}{Zeljko Carevic}, \bibinfo{person}{Maria
  Lusky}, \bibinfo{person}{Wilko van Hoek}, {and} \bibinfo{person}{Philipp
  Mayr}.} \bibinfo{year}{2017}\natexlab{}.
\newblock \showarticletitle{Investigating exploratory search activities based
  on the stratagem level in digital libraries}.
\newblock \bibinfo{journal}{\emph{International Journal on Digital Libraries}}
  (\bibinfo{year}{2017}), \bibinfo{pages}{1--21}.
\newblock
\urldef\tempurl%
\url{http://link.springer.com/10.1007/s00799-017-0226-6}
\showURL{%
\tempurl}


\bibitem[\protect\citeauthoryear{Carevic and Mayr}{Carevic and Mayr}{2015}]%
        {carevic2015extending}
\bibfield{author}{\bibinfo{person}{Zeljko Carevic} {and}
  \bibinfo{person}{Philipp Mayr}.} \bibinfo{year}{2015}\natexlab{}.
\newblock \showarticletitle{{Extending search facilities via
  bibliometric-enhanced stratagems}}.
\newblock  (\bibinfo{year}{2015}), \bibinfo{pages}{40--46}.
\newblock
\urldef\tempurl%
\url{http://ceur-ws.org/Vol-1344/paper5.pdf}
\showURL{%
\tempurl}


\bibitem[\protect\citeauthoryear{Carevic and Mayr}{Carevic and Mayr}{2016}]%
        {carevic2016survey}
\bibfield{author}{\bibinfo{person}{Zeljko Carevic} {and}
  \bibinfo{person}{Philipp Mayr}.} \bibinfo{year}{2016}\natexlab{}.
\newblock \showarticletitle{Survey on High-level Search Activities based on the
  Stratagem Level in Digital Libraries}. In
  \bibinfo{booktitle}{\emph{International Conference on Theory and Practice of
  Digital Libraries}}. Springer, \bibinfo{pages}{54--66}.
\newblock
\urldef\tempurl%
\url{http://link.springer.com/10.1007/978-3-319-43997-6\_5}
\showURL{%
\tempurl}


\bibitem[\protect\citeauthoryear{Ellis}{Ellis}{1989}]%
        {ellis1989behavioural}
\bibfield{author}{\bibinfo{person}{David Ellis}.}
  \bibinfo{year}{1989}\natexlab{}.
\newblock \showarticletitle{A behavioural approach to information retrieval
  system design}.
\newblock \bibinfo{journal}{\emph{Journal of Documentation}}
  \bibinfo{volume}{45}, \bibinfo{number}{3} (\bibinfo{year}{1989}),
  \bibinfo{pages}{171--212}.
\newblock


\bibitem[\protect\citeauthoryear{Fagan}{Fagan}{2010}]%
        {fagan2010usability}
\bibfield{author}{\bibinfo{person}{Jody~Condit Fagan}.}
  \bibinfo{year}{2010}\natexlab{}.
\newblock \showarticletitle{Usability studies of faceted browsing: A literature
  review}.
\newblock \bibinfo{journal}{\emph{Information Technology and Libraries}}
  \bibinfo{volume}{29}, \bibinfo{number}{2} (\bibinfo{year}{2010}),
  \bibinfo{pages}{58}.
\newblock


\bibitem[\protect\citeauthoryear{Frias-Martinez, Chen, and Liu}{Frias-Martinez
  et~al\mbox{.}}{2009}]%
        {Frias-Martinez:2009}
\bibfield{author}{\bibinfo{person}{Enrique Frias-Martinez},
  \bibinfo{person}{Sherry~Y. Chen}, {and} \bibinfo{person}{Xiaohui Liu}.}
  \bibinfo{year}{2009}\natexlab{}.
\newblock \showarticletitle{Evaluation of a Personalized Digital Library Based
  on Cognitive Styles: Adaptivity vs. Adaptability}.
\newblock \bibinfo{journal}{\emph{Int. J. Inf. Manag.}} \bibinfo{volume}{29},
  \bibinfo{number}{1} (\bibinfo{date}{Feb.} \bibinfo{year}{2009}),
  \bibinfo{pages}{48--56}.
\newblock
\showISSN{0268-4012}


\bibitem[\protect\citeauthoryear{Fuhr}{Fuhr}{2017}]%
        {Fuhr:17b}
\bibfield{author}{\bibinfo{person}{Norbert Fuhr}.}
  \bibinfo{year}{2017}\natexlab{}.
\newblock \bibinfo{booktitle}{\emph{Some Common Mistakes In IR Evaluation, And
  How They Can Be Avoided}}.
\newblock \bibinfo{type}{{T}echnical {R}eport}.
  \bibinfo{institution}{University of Duisburg-Essen, Germany}.
\newblock


\bibitem[\protect\citeauthoryear{Hearst}{Hearst}{2009}]%
        {hearst2009search}
\bibfield{author}{\bibinfo{person}{Marti Hearst}.}
  \bibinfo{year}{2009}\natexlab{}.
\newblock \bibinfo{booktitle}{\emph{Search user interfaces}}.
\newblock \bibinfo{publisher}{Cambridge University Press}.
\newblock


\bibitem[\protect\citeauthoryear{Hearst}{Hearst}{2006}]%
        {hearst2006clustering}
\bibfield{author}{\bibinfo{person}{Marti~A Hearst}.}
  \bibinfo{year}{2006}\natexlab{}.
\newblock \showarticletitle{Clustering versus faceted categories for
  information exploration}.
\newblock \bibinfo{journal}{\emph{Commun. ACM}} \bibinfo{volume}{49},
  \bibinfo{number}{4} (\bibinfo{year}{2006}), \bibinfo{pages}{59--61}.
\newblock


\bibitem[\protect\citeauthoryear{Hienert and Mutschke}{Hienert and
  Mutschke}{2016}]%
        {hienert2016usefulness}
\bibfield{author}{\bibinfo{person}{Daniel Hienert} {and} \bibinfo{person}{Peter
  Mutschke}.} \bibinfo{year}{2016}\natexlab{}.
\newblock \showarticletitle{A usefulness-based approach for measuring the local
  and global effect of IIR services}. In \bibinfo{booktitle}{\emph{Proceedings
  of the 2016 ACM Conference on Human Information Interaction and Retrieval}}.
  ACM, \bibinfo{pages}{153--162}.
\newblock


\bibitem[\protect\citeauthoryear{Hienert, Sawitzki, and Mayr}{Hienert
  et~al\mbox{.}}{2015}]%
        {Hienert2015}
\bibfield{author}{\bibinfo{person}{Daniel Hienert}, \bibinfo{person}{Frank
  Sawitzki}, {and} \bibinfo{person}{Philipp Mayr}.}
  \bibinfo{year}{2015}\natexlab{}.
\newblock \showarticletitle{Digital library research in action: supporting
  information retrieval in Sowiport}.
\newblock \bibinfo{journal}{\emph{D-Lib Magazine}} \bibinfo{volume}{21},
  \bibinfo{number}{3} (\bibinfo{year}{2015}), \bibinfo{pages}{8}.
\newblock
\urldef\tempurl%
\url{https://doi.org/10.1045/march2015-hienert}
\showDOI{\tempurl}


\bibitem[\protect\citeauthoryear{Joachims et~al\mbox{.}}{Joachims
  et~al\mbox{.}}{2003}]%
        {joachims2003evaluating}
\bibfield{author}{\bibinfo{person}{Thorsten Joachims} {et~al\mbox{.}}}
  \bibinfo{year}{2003}\natexlab{}.
\newblock \bibinfo{title}{Evaluating Retrieval Performance Using Clickthrough
  Data.}
\newblock   (\bibinfo{year}{2003}).
\newblock


\bibitem[\protect\citeauthoryear{Kanoulas, Carterette, Hall, Clough, and
  Sanderson}{Kanoulas et~al\mbox{.}}{2011}]%
        {Kanuoulas/etal:11}
\bibfield{author}{\bibinfo{person}{Evangelos Kanoulas}, \bibinfo{person}{Ben
  Carterette}, \bibinfo{person}{Mark Hall}, \bibinfo{person}{Paul Clough},
  {and} \bibinfo{person}{Mark Sanderson}.} \bibinfo{year}{2011}\natexlab{}.
\newblock \showarticletitle{Overview of the TREC 2011 Session Track}. In
  \bibinfo{booktitle}{\emph{TREC 2011}}. \bibinfo{publisher}{NIST}.
\newblock


\bibitem[\protect\citeauthoryear{Lagoze, Kraft, Payette, and Jesuroga}{Lagoze
  et~al\mbox{.}}{2005}]%
        {lagoze2005}
\bibfield{author}{\bibinfo{person}{Carl Lagoze}, \bibinfo{person}{Dean~B.
  Kraft}, \bibinfo{person}{Sandy Payette}, {and} \bibinfo{person}{Susan
  Jesuroga}.} \bibinfo{year}{2005}\natexlab{}.
\newblock \showarticletitle{What {Is} a {Digital} {Library} {Anyway}?: {Beyond}
  {Search} and {Access} in the {NSDL}}.
\newblock \bibinfo{journal}{\emph{D-Lib Magazine}} \bibinfo{volume}{11},
  \bibinfo{number}{11} (\bibinfo{date}{Nov.} \bibinfo{year}{2005}).
\newblock
\showISSN{1082-9873}
\urldef\tempurl%
\url{https://doi.org/10.1045/november2005-lagoze}
\showDOI{\tempurl}


\bibitem[\protect\citeauthoryear{Marchionini}{Marchionini}{2006}]%
        {marchionini2006exploratory}
\bibfield{author}{\bibinfo{person}{Gary Marchionini}.}
  \bibinfo{year}{2006}\natexlab{}.
\newblock \showarticletitle{Exploratory search: from finding to understanding}.
\newblock \bibinfo{journal}{\emph{Commun. ACM}} \bibinfo{volume}{49},
  \bibinfo{number}{4} (\bibinfo{year}{2006}), \bibinfo{pages}{41--46}.
\newblock


\bibitem[\protect\citeauthoryear{Matthijs and Radlinski}{Matthijs and
  Radlinski}{2011}]%
        {matthijs2011personalizing}
\bibfield{author}{\bibinfo{person}{Nicolaas Matthijs} {and}
  \bibinfo{person}{Filip Radlinski}.} \bibinfo{year}{2011}\natexlab{}.
\newblock \showarticletitle{Personalizing web search using long term browsing
  history}. In \bibinfo{booktitle}{\emph{Proceedings of the fourth ACM
  international conference on Web search and data mining}}. ACM,
  \bibinfo{pages}{25--34}.
\newblock


\bibitem[\protect\citeauthoryear{Mayr and Kacem}{Mayr and Kacem}{2017}]%
        {mayr2017complete}
\bibfield{author}{\bibinfo{person}{Philipp Mayr} {and} \bibinfo{person}{Ameni
  Kacem}.} \bibinfo{year}{2017}\natexlab{}.
\newblock \showarticletitle{{A Complete Year of User Retrieval Sessions in a
  Social Sciences Academic Search Engine}}.
\newblock  (\bibinfo{year}{2017}), \bibinfo{pages}{560--565}.
\newblock
\urldef\tempurl%
\url{https://doi.org/10.1007/978-3-319-67008-9_46}
\showDOI{\tempurl}


\bibitem[\protect\citeauthoryear{Mayr and Petras}{Mayr and Petras}{2008}]%
        {mayr2008}
\bibfield{author}{\bibinfo{person}{Philipp Mayr} {and} \bibinfo{person}{Vivien
  Petras}.} \bibinfo{year}{2008}\natexlab{}.
\newblock \showarticletitle{Cross-concordances: terminology mapping and its
  effectiveness for information retrieval}. In \bibinfo{booktitle}{\emph{74th
  {IFLA} {World} {Library} and {Information} {Congress}}}.
  \bibinfo{publisher}{IFLA}, \bibinfo{address}{Quebec, Canada}.
\newblock
\urldef\tempurl%
\url{http://www.ifla.org/IV/ifla74/papers/129-Mayr_Petras-en.pdf}
\showURL{%
\tempurl}


\bibitem[\protect\citeauthoryear{Pharo, Beckers, Nordlie, and Fuhr}{Pharo
  et~al\mbox{.}}{2011}]%
        {Pharo/etal:11a}
\bibfield{author}{\bibinfo{person}{Nils Pharo}, \bibinfo{person}{Thomas
  Beckers}, \bibinfo{person}{Ragnar Nordlie}, {and} \bibinfo{person}{Norbert
  Fuhr}.} \bibinfo{year}{2011}\natexlab{}.
\newblock \showarticletitle{Overview of the INEX 2010 Interactive Track}.
  \bibinfo{pages}{227--235}.
\newblock


\bibitem[\protect\citeauthoryear{Renda and Straccia}{Renda and
  Straccia}{2005}]%
        {renda2005}
\bibfield{author}{\bibinfo{person}{M.Elena Renda} {and}
  \bibinfo{person}{Umberto Straccia}.} \bibinfo{year}{2005}\natexlab{}.
\newblock \showarticletitle{A personalized collaborative {Digital} {Library}
  environment: a model and an application}.
\newblock \bibinfo{journal}{\emph{Information Processing \& Management}}
  \bibinfo{volume}{41}, \bibinfo{number}{1} (\bibinfo{date}{Jan.}
  \bibinfo{year}{2005}), \bibinfo{pages}{5--21}.
\newblock
\showISSN{03064573}
\urldef\tempurl%
\url{https://doi.org/10.1016/j.ipm.2004.04.007}
\showDOI{\tempurl}


\bibitem[\protect\citeauthoryear{Ruthven, Lalmas, and Van~Rijsbergen}{Ruthven
  et~al\mbox{.}}{2003}]%
        {ruthven2003incorporating}
\bibfield{author}{\bibinfo{person}{Ian Ruthven}, \bibinfo{person}{Mounia
  Lalmas}, {and} \bibinfo{person}{Keith Van~Rijsbergen}.}
  \bibinfo{year}{2003}\natexlab{}.
\newblock \showarticletitle{Incorporating user search behavior into relevance
  feedback}.
\newblock \bibinfo{journal}{\emph{Journal of the Association for Information
  Science and Technology}} \bibinfo{volume}{54}, \bibinfo{number}{6}
  (\bibinfo{year}{2003}), \bibinfo{pages}{529--549}.
\newblock


\bibitem[\protect\citeauthoryear{Sepliarskaia, Radlinski, and
  de~Rijke}{Sepliarskaia et~al\mbox{.}}{2017}]%
        {sepliarskaia2017simple}
\bibfield{author}{\bibinfo{person}{Anna Sepliarskaia}, \bibinfo{person}{Filip
  Radlinski}, {and} \bibinfo{person}{Maarten de Rijke}.}
  \bibinfo{year}{2017}\natexlab{}.
\newblock \showarticletitle{Simple Personalized Search Based on Long-Term
  Behavioral Signals}. In \bibinfo{booktitle}{\emph{European Conference on
  Information Retrieval}}. Springer, \bibinfo{pages}{95--107}.
\newblock


\bibitem[\protect\citeauthoryear{Shen, Tan, and Zhai}{Shen
  et~al\mbox{.}}{2005}]%
        {shen2005context}
\bibfield{author}{\bibinfo{person}{Xuehua Shen}, \bibinfo{person}{Bin Tan},
  {and} \bibinfo{person}{ChengXiang Zhai}.} \bibinfo{year}{2005}\natexlab{}.
\newblock \showarticletitle{Context-sensitive information retrieval using
  implicit feedback}. In \bibinfo{booktitle}{\emph{28th SIGIR conference}}.
  ACM, \bibinfo{pages}{43--50}.
\newblock


\bibitem[\protect\citeauthoryear{Smeaton and Callan}{Smeaton and
  Callan}{2005}]%
        {Smeaton2005}
\bibfield{author}{\bibinfo{person}{Alan~F. Smeaton} {and}
  \bibinfo{person}{Jamie Callan}.} \bibinfo{year}{2005}\natexlab{}.
\newblock \showarticletitle{Personalisation and recommender systems in digital
  libraries}.
\newblock \bibinfo{journal}{\emph{International Journal on Digital Libraries}}
  \bibinfo{volume}{5}, \bibinfo{number}{4} (\bibinfo{date}{01 Aug}
  \bibinfo{year}{2005}), \bibinfo{pages}{299--308}.
\newblock
\showISSN{1432-1300}
\urldef\tempurl%
\url{https://doi.org/10.1007/s00799-004-0100-1}
\showDOI{\tempurl}


\bibitem[\protect\citeauthoryear{Teevan, Liebling, and
  Ravichandran~Geetha}{Teevan et~al\mbox{.}}{2011}]%
        {teevan2011understanding}
\bibfield{author}{\bibinfo{person}{Jaime Teevan}, \bibinfo{person}{Daniel~J
  Liebling}, {and} \bibinfo{person}{Gayathri Ravichandran~Geetha}.}
  \bibinfo{year}{2011}\natexlab{}.
\newblock \showarticletitle{Understanding and predicting personal navigation}.
  In \bibinfo{booktitle}{\emph{Proceedings of the fourth ACM international
  conference on Web search and data mining}}. ACM, \bibinfo{pages}{85--94}.
\newblock


\bibitem[\protect\citeauthoryear{White, Jose, van Rijsbergen, and
  Ruthven}{White et~al\mbox{.}}{2004}]%
        {white2004simulated}
\bibfield{author}{\bibinfo{person}{Ryen White}, \bibinfo{person}{Joemon Jose},
  \bibinfo{person}{C van Rijsbergen}, {and} \bibinfo{person}{Ian Ruthven}.}
  \bibinfo{year}{2004}\natexlab{}.
\newblock \showarticletitle{A simulated study of implicit feedback models}.
\newblock \bibinfo{journal}{\emph{Advances in Information Retrieval}}
  (\bibinfo{year}{2004}), \bibinfo{pages}{311--326}.
\newblock


\bibitem[\protect\citeauthoryear{White, Chu, Hassan, He, Song, and Wang}{White
  et~al\mbox{.}}{2013}]%
        {white2013enhancing}
\bibfield{author}{\bibinfo{person}{Ryen~W White}, \bibinfo{person}{Wei Chu},
  \bibinfo{person}{Ahmed Hassan}, \bibinfo{person}{Xiaodong He},
  \bibinfo{person}{Yang Song}, {and} \bibinfo{person}{Hongning Wang}.}
  \bibinfo{year}{2013}\natexlab{}.
\newblock \showarticletitle{Enhancing personalized search by mining and
  modeling task behavior}. In \bibinfo{booktitle}{\emph{22nd international
  conference on World Wide Web}}. ACM, \bibinfo{pages}{1411--1420}.
\newblock


\bibitem[\protect\citeauthoryear{White and Drucker}{White and Drucker}{2007}]%
        {white2007investigating}
\bibfield{author}{\bibinfo{person}{Ryen~W White} {and}
  \bibinfo{person}{Steven~M Drucker}.} \bibinfo{year}{2007}\natexlab{}.
\newblock \showarticletitle{Investigating behavioral variability in web
  search}. In \bibinfo{booktitle}{\emph{Proceedings of the 16th international
  conference on World Wide Web}}. ACM, \bibinfo{pages}{21--30}.
\newblock


\bibitem[\protect\citeauthoryear{White, Jose, and Ruthven}{White
  et~al\mbox{.}}{2002}]%
        {white2002comparing}
\bibfield{author}{\bibinfo{person}{Ryen~W White}, \bibinfo{person}{Joemon~M
  Jose}, {and} \bibinfo{person}{Ian Ruthven}.} \bibinfo{year}{2002}\natexlab{}.
\newblock \showarticletitle{Comparing explicit and implicit feedback techniques
  for web retrieval: Trec-10 interactive track report}. In
  \bibinfo{booktitle}{\emph{Proceedings of the Tenth Text Retrieval Conference
  (TREC-10)}}. \bibinfo{pages}{534--538}.
\newblock


\end{thebibliography}
